# Benchmark Studies on the Isomerization Enthalpies for Interstellar Molecular Species

Emmanuel E. Etim
Department of Chemical Sciences, Federal University Wukari, Katsina-Ala Road, P.M.B. 1020 Wukari, Taraba State, Nigeria
Department of Chemistry, Indian Institute of Technology Bombay, Powai, Mumbai 400 076, India
*email: emmaetim@gmail.com

**Abstract:** With the well-established correlation between the relative stabilities of isomers and their interstellar abundances coupled with the prevalence of isomeric species among the interstellar molecular species, isomerization remains a plausible formation route for isomers in the interstellar medium. The present work reports an extensive investigation of the isomerization energies of 246 molecular species from 65 isomeric groups using the Gaussian-4 theory composite method with atoms ranging from 3 to 12. From the results, the high abundances of the most stable isomers coupled with the energy sources in interstellar medium drive the isomerization process even for barriers as high as 67.4 kcal/mol. Specifically, the cyanides and their corresponding isocyanides pairs appear to be effectively synthesized via this process. The following potential interstellar molecules; CNC, NCCN, c-C5H, methylene ketene, methyl Ketene, CH3SCH3, C5O, 1,1-ethanediol, propanoic acid, propan-2-ol, and propanol are identified and discussed. The study further reaffirms the importance of thermodynamics in interstellar formation processes on a larger scale and accounts for the known isomeric species. In all the isomeric groups, isomerization appears to be an effective route for the formation of the less stable isomers (which are probably less abundant) from the most stable ones that are perhaps more abundant.

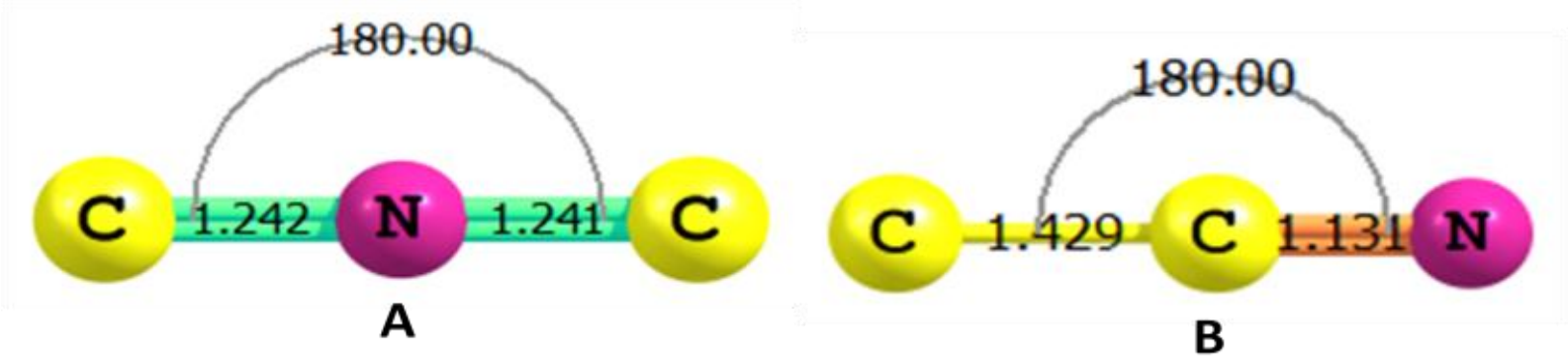

**Introduction**: That the interstellar medium (ISM) is chemically rich is not an argument with the discoveries of over 200 different molecular species in this thin space between the stars which was earlier regarded as a vacuum dotted with stars, black holes, and other celestial bodies [1-6]. These molecules are important in various fields such as atmospheric chemistry, astrochemistry, prebiotic chemistry, astrophysics, astronomy, astrobiology, etc, and in our understanding of the solar system, "the world around us" with each successfully detected interstellar molecule telling the story of the chemistry and physics of the environment from where it was detected. They serve as the most important tools for probing deep into the interior of the molecular clouds and the molecular clouds are significant because it is from them that stars and consequently new planets are formed. The symmetric rotors serve as interstellar thermometers while the metal-bearing species provide useful information regarding the depletion of these molecular species into the molecular dust grains. Understanding how the simple molecules that were present on the early earth may have given rise to the complex systems and processes of contemporary biology is one question the biologically related interstellar molecules can be used to address. Molecules also provide the cooling mechanism for the clouds through their emission [3,7-8].

A careful look at the list of known interstellar and circumstellar molecules reveals some interesting chemistries among these molecular species. Tables 1 and 2 contain the list of the currently (as of June 2021) known interstellar and circumstellar species arranged according to the number of atoms making up the molecules.

Table 1: Interstellar molecules between 2 and 7 atoms

| 2 atoms | 3 atoms | 4 atoms | 5 atoms | 6 atoms | 7 atoms |
|---|---|---|---|---|---|
| $H_2$, CO, CSi, CP<br>CS, NO, NS, SO<br>HCl, NaCl, KCl<br>AlCl, AlF, PN<br>SiN, SiO, SiS<br>NH, OH, $C_2$<br>CN, HF, FeO<br>LiH, CH, $CH^+$<br>$CO^+$, $SO^+$, SH, $NO^+$<br>$O_2$, $N_2$, $CF^+$<br>PO, HD<br>SiH, AlO,<br>$ArH^+$ , $OH^+$, $CN^-$, $SH^+$, $HCl^+$, TiO, CrO, $NS^+$,<br>VO, $HeH^+$ | $H_2O$, $H_2S$, HCN, $TiO_2$, HNC, $CO_2$, $SO_2$<br>MgCN, MgNC, NaCN , FeCN, KCN,<br>AlOH, $H_2Cl$, $H_2O^+$, $H_2Cl^+$, $N_2O$, $NH_2$, OCS<br>$CH_2$, HCO, $C_3$<br>$C_2H$, $C_2O$, $C_2S$ , AlNC, HNO<br>SiCN, $N_2H^+$ SiNC, c-$SiC_2$<br>$C_2N$, HCO+, $HOC^+$ $HCS^+$, $H_3^+$ $OCN^-$, HCP, CCP, SiCSi , $S_2H$, HCS, HSC, NCO,<br>CaNC,<br>NCS, $HO_2$ | $NH_3$, $H_2CO$<br>$H_2CS$, $C_2H_2$<br>HNCS<br>$H_3O^+$, $SiC_3$<br>$C_3S$, $H_2CN$<br>c-$C_3H$, l-$C_3H$<br>HCCN, $CH_3$<br>$C_2CN$, $C_3O$<br>$HCNH^+$, $HOCO^+$<br>$C_3N^-$, HNCO<br>HSCN, $C_3N$, $PH_3$<br>HMgNC<br>HCCO, NCCP, MgCCH, HOCN,<br>CNCN<br>$H_2O_2$ ,<br>t-HONO | $CH_4$, $SiH_4$<br>$CH_2NH$, $NH_2CN$<br>$CH_2CO$, HCOOH<br>HCCCN, HCCNC<br>c-$C_3H_2$, l-$C_3H_2$<br>$CH_2CN$, $H_2COH^+$<br>$C_4Si$<br>$C_5$<br>HNCCC<br>$C_4H$<br>$C_4H^-$<br>HC(O)CN, $CH_3O$, HNCNH, $H_2NCO^+$, $NH_3D^+$, $CH_3Cl$,<br>$MgC_3N$, $NH_2OH$,<br>$HC_3O^+$,<br>$HC_3S^+$,<br>$C_4S$, | $CH_3OH$, $CH_3SH$<br>$C_2H_4$, $HC_4H$<br>$CH_3CN$, $CH_3NC$<br>$HCONH_2$,<br>$HC_2C(O)H$<br>$HC_3NH^+$<br>$HC_4N$<br>$C_5N$, $C_5H$<br>$H_2C_4$, $H_2CCNH$<br>$C_5N^-$<br>c-$H_2C_3O$<br>HNCHCN, $C_5S$, $SiH_3CN$,<br>z-HNCHCN,<br>$MgC_4H$<br>$CH_3CO^+$,<br>$CH_2CCH$,<br>$H_2CCCS$, | $CH_2CHOH$<br>c-$C_2H_4O$<br>$HC(O)CH_3$<br>$H_3CCCH$<br>$CH_3NH_2$<br>$CH_2CHCN$<br>$HC_4CN$<br>$C_6H$<br>$CH_3NCO$<br>$HC_5O$,<br>$HOCH_2CN$<br>HNCHCCH,<br>$HC_4NC$,<br>c-$C_3HCCH$, |
| | | | **t-HC(O)SH,**<br>**$H_2CCS$,**<br>**$NCCNH^+$** | | |

Table 2: Interstellar molecules between 8 and >12 atoms

| **8 atoms** | **9 atoms** | **10 atoms** | **11 atoms** | **12 atoms** | **>12 atoms** |
|---|---|---|---|---|---|
| **$CH_3COOH$** | **$(CH_3)_2O$** | **$(CH_3)CO$** | **$HC_9N$** | **$C_6H_6$** | **$HC_{11}N$** |
| **$HCOOCH_3$** | **$CH_3CH_2CN$** | **$HOC_2H_4OH$** | **$CH_3C_6H$** | **$C_3H_7CN$** | **$C_{60}$** |
| **$HOCH_2CHO$** | **$CH_3CH_2OH$** | **$H_3CCH_2COH$** | **$HCOOC_2H_5$** | **$C_2H_5OCH_3$** | **$C_{60}^+$** |
| **$H_3C_3CN$,**<br>**$C_7H$,**<br>**$(NH_2)_2CO$** | **$CH_3CH_2SH$** | **$CH_3C_4CN$,**<br>**$CH_3CHCHO$,** | **$CH_3OCOCH_3$** | **branched-$C_3H_7CN$ ,** | **$C_{70}$** |
| | **$CH_3C_4H$** | **$CH_3OCH_2OH$** | **$CH_3COCH_2OH$** | **c-$C_5H_5CN$** | **c-$C_6H_5CN$** |
| **$H_2C_6$** | **$HC_7N$** | | **c-$C_5H_6$** | | **$C_{10}H_7CN$** |
| **$H(CC)_3H$** | **$C_8H$** | | **$NH_2CH_2CH_2OH$** | | **c-$C_9H_8$** |
| **$H_2CCHCHO$** | **$CH_3CONH_2$** | | | | |
| **$CH_2CCHCN$** | **$C_8H^-$** | | | | |
| **$H_2NCH_2CN$,** | **$CH_3CHCH_2$,** | | | | |
| **$CH_3CHNH$** | **$CH_3NHCHO$,** | | | | |
| **$CH_3SiH_3$,** | **$HC_7O$** | | | | |
| **$(NH_2)_2CO$** | **HCCCHCHCN** | | | | |
| **$HCCCH_2CN$,** | **$H_2CCHC_3N$** | | | | |
| **$HC_5NH^+$,** | **$H_2CCCHCCH$** | | | | |
| **$CH_2CHCCH$,** | | | | | |

Some of the chemistries existing among the interstellar molecular species include the dominance of the linear carbon chain species of the form $C_n$, $H_2C_n$, $HC_nN$, $CH_3(CC)_nH$, $CH_3(C{\equiv}C)_nCN$ and $C_nX$ (X=N, O, Si, S, H, P, $N^-$, $H^-$) which account for more than 20% of all the known interstellar and circumstellar molecular species; the presence of about 30 alkyl group containing interstellar molecules mostly observed from the same or similar molecular clouds; periodic trends in which elements from the same group appear to form similar compounds with similar properties as in the case of oxygen and sulphur (group 6/16 elements) in which of the over 25 S-containing molecules observed in ISM, about 20 have the corresponding O-containing molecules uniquely detected in ISM and the abundance of S-compound relative to its O-analogue is approximately equal to the cosmic S/O ratio, 1/42 as seen in methyl mercaptan, thioisocyanic acid, etc, except where the effect of interstellar hydrogen bonding dominates [9-10]; successive hydrogen addition where larger species are believed to be formed from the smaller unsaturated species via successive hydrogen addition in which both species could be shown to be chemically and spatially related [11]. Isomerism is yet another prevailing chemistry among the interstellar molecular species. Apart from the diatomics and a few species which cannot form isomers, about 40% of all the known interstellar molecules have isomeric analogues ranging from the isomeric pairs to the isomeric triads which are believed to have a common precursor for their formation process [12]. Table 3 lists some of the known isomeric species and their isomeric groups.

Table 3: Known interstellar isomeric pairs, triads and their isomeric groups.

| **Isomeric Group** | **Astronomically observed isomers** |
|---|---|
| CHN | HCN, HNC |
| CNMg | MgCN, MgNC |
| CNSi | SiCN, SiNC |
| $CHO^+$ | $HOC^+$, $HCO^+$ |
| CHSN | HSCN, HNCS |
| $HC_3$ | c-$C_3H$, l-$C_3H$ |
| $H_2C_3$ | c-$C_3H_2$, l-$C_3H_2$ |
| $CN_2H_2$ | $NH_2CN$, HNCNH |
| $C_3H_2O$ | HCCCHO, c-$H_2C_3O$ |
| $C_4H_2$ | $H(CC)_2H$, $H_2C_4$ |
| $C_6H_2$ | $H_2C_6$, $H(CC)_3H$ |
| $C_4H_3N$ | $CH_2CCHCN$, $CH_3CCCN$ |
| $C_2H_6O$ | $(CH_3)_2O$, $CH_3CH_2OH$ |
| $C_3H_6O$ | $(CH_3)_2CO$, $CH_3CH_2C(O)H$ |
| $C_3H_6O_2$ | $HC(O)OCH_2CH_3$, $CH_3OC(O)CH_3$ |
| $C_4H_7N$ | n-$CH_3CH_2CH_2CN$, b-$CH_3CH_2CH_2CN$ |
| CHON | HNCO, HCNO, HOCN |
| $C_2H_4O_2$ | $CH_3COOH$, $HC(O)OCH_3$, $HOCH_2C(O)H$ |
| $C_2H_4O$ | $CH_2CH(OH)$ c-$C_2H_4O$, $HC(O)CH_3$ |
| $C_2H_3N$ | $CH_3CN$, $CH_3NC$, $H_2CCNH$ |
| $C_3HN$ | $HC_2CN$, $HC_2NC$ HNCCC |

Because of the conditions (low temperature and pressure) in the interstellar medium, there is hardly a consensus as to how these molecules are formed but some of the chemistries listed above and the effect of thermodynamics serve as clues as to how these molecules could be formed in ISM. Focusing here on isomerization, the prevalence of isomeric species among the known astromolecules coupled with the energy sources in ISM such as the shock waves (which could arise from the interaction of the Earth's magnetic field with the solar wind, molecular outflows during star formation, supernova blasts and galaxies colliding with each other) which provide energy for both the formation and distribution of large interstellar species, place isomerization as one of the most plausible routes for the formation of interstellar molecules [13]. As observed in some studies, the most stable isomers are found to be more abundant than their less stable analogues except where other factors dominate; thus, the isomerization of the most stable isomer (which is probably the most abundant) to the less stable isomers can be a very effective and efficient formation mechanism in ISM. Also, apart from the energy sources in ISM, the high abundance of the most stable isomer can drive the isomerization process irrespective of the energy barrier.

According to the minimum energy principle [14], isomerization is the most important process in determining the relative abundances of the isomers in ISM. Isomers of the same generic formula are said to have a common intermediate in their formation and destruction routes. After reaching the generic formula, the equilibration process is said to occur. This implies an internal isomerization with a low activation barrier, assisted isomerization, or catalytic isomerization at the grain/ice surface [14].

The present work aims at estimating accurate isomerization enthalpies for 246 different molecular species from 65 isomeric groups using the Gaussian 4 theory composite method [15]. The molecules range from the 3 atomic species to those with 12 atoms with at least one known interstellar molecule from each isomeric group. The results account for the extent and effectiveness of isomerization as a plausible formation route in ISM; and the rationale behind the successful observation of the known species. Among other things, potential candidates for astronomical searches are highlighted and discussed.

**Computational Details**: A concerted effort between theory and experiment is found in the heart of some of the most successful scientific studies. Considering the large range of molecules examined in this study, only very few of them have experimentally measured enthalpies of formation while others are so unstable that they cannot be probed experimentally in the terrestrial laboratory but all of these can be comfortably handled computationally. The Gaussian 4 (G4) theory composite method is employed in estimating accurate standard enthalpies of formation ($\Delta_f H^O$) for all the molecular systems investigated in this work. The G4 composite method is very accurate for several systems and benchmark studies; in its release note, it has an average absolute deviation of 0.8 kcal/mol from experimental values for the enthalpies of formation of 270 molecular species. The G4 method is a modification of the G3 method which has an average absolute deviation of 1.19 kcal/mol for the same number of molecular systems [15-16]. In the G4 method, geometry optimization and zero point energy (ZPE) are carried out at the B3LYP level of theory using the 6-31G(2df,p) basis set. The ZPE is scaled by 0.9854. Single-point calculations and energy are done using the MP4/6-31G(d) method modified by corrections from additional calculations (with MP4 and other methods) while core correlation is obtained via higher-level correction terms. The Gaussian 09 suite of programs is used for all the computational studies reported here. Only stable geometries with no imaginary frequencies are considered. The standard enthalpies of formation are obtained from the atomization energies using the approach described in previous studies [4-6, 17-23]

## Results and Discussion

The standard enthalpies of formation ($\Delta_f H^O$) for all the 246 molecular species from 65 isomeric groups examined in this study are contained in the supporting information. The relative enthalpies for each isomeric group are presented and discussed in this section. The different isomeric groups investigated are grouped according to the number of atoms beginning from 3 to 12.

### Isomers with 3 atoms

In Table 4, the relative enthalpies or isomerization energies of the 26 molecular species from 13 isomeric groups with 3 atoms are shown alongside the current astronomical status of these species. At least one isomer from each of these isomeric groups is a known interstellar molecule [24-43]. The isomerization enthalpies range from 0.2 to 41.5 kcal/mol for groups where both isomers have been detected. This range illustrates the effectiveness of the isomerization mechanism as a plausible route for the formation of these molecular species in ISM; it also shows how far or the extent to which the energy sources within the ISM can drive some chemical processes. For the isomeric groups where only one isomer has been detected, the observed range of the relative enthalpy of formation suggests that species like NaNC, AlCN, $ONC^-$ and even HOC can be formed from their most stable isomers that have been detected already. The high abundances of the stable isomers coupled with the energy sources in ISM imply the possibility of these less stable isomers being formed from their most stable isomers via the isomerization mechanism.

With the exception of $C_2N$ group where only the less stable isomer is been detected, in all other cases where only one isomer is detected, it is the most stable isomer which supports the fact that the most stable isomer is probably the most abundant and the most abundant species is easily detected compared to the less stable isomer. Where both isomers have been detected, the most stable isomer is found to be the most abundant except where other processes dominate. CNC is more stable than CCN but the less stable isomer has been detected while the most stable isomer is yet to be astronomically observed. From literature perusal, there is no information regarding the spectroscopic parameters of CNC that would have warranted it astronomical searches. Thus, the detection of CNC awaits the availability of accurate spectroscopic parameters.

Ions (both cations and anions) play important role in the formation processes of interstellar molecules. Under the conditions of the ISM, neutral atoms and molecules tend to be unreactive toward molecular hydrogen but in the presence of ions, most of the reactions become very efficient since the ions can easily react without having to overcome the reaction barrier [7,44]. This can be seen in the case of the $CHO^+$ group where $HOC^+$ with a relative enthalpy of formation of 37.3 kcal/mol has been detected. Theoretical calculations and laboratory experiments have shown that the isomerization process in the $CHO^+$ isomers has essentially no barrier [14]. Thus, with the availability of accurate spectroscopic parameters, the $HSC^+$ and $ONC^-$ ions (where their most stable isomers are already detected) could be successfully detected.

Isomerization appears to be a favourable route for the cyanide/isocyanide pair. Table 4 contains 7 cyanide/isocyanide pairs of which both cyanide and isocyanide have been detected in three groups with isomerization energy ranging from 0.2 to 13.5 kcal/mol. Except for the $ONC^-$ ion with a relative enthalpy of formation of 17.1kcal/mol; the remaining members of the cyanide/isocyanide pairs have relative enthalpy of formation in the range of 0.0 to 7.3 kcal/mol which is well within the range of those already detected. Thus, NaNC, AlCN and CNC are most likely to be formed from their corresponding analogues and could be successfully detected.

Table 4: Isomerization enthalpies for isomers with 3 atoms

| Isomeric group | Isomers | Relative $\Delta_f H^O$ (kcal/mol) | Astronomical status | Ref. |
|---|---|---|---|---|
| CNH | HCN | 0.0 | Observed | [24] |
| | HNC | 13.5 | Observed | [25] |
| | | | | |
| CNNa | NaCN | 0.0 | Observed | [26] |
| | NaNC | 2.5 | Not observed | |
| | | | | |
| CNMg | MgNC | 0.0 | Observed | [27] |
| | MgCN | 0.7 | Observed | [28-30] |
| | | | | |
| CNAl | AlNC | 0.0 | Observed | [31] |
| | AlCN | 7.3 | Not observed | |
| | | | | |
| CNSi | SiNC | 0.0 | Observed | [32] |
| | SiCN | 0.2 | Observed | [33] |
| | | | | |
| CHO | HCO | 0.0 | Observed | [34] |
| | HOC | 38.2 | Not observed | |
| | | | | |
| CHP | HCP | 0.0 | Observed | [35] |
| | HPC | 75.7 | Not observed | |
| | | | | |
| $C_2N$ | CNC | 0.0 | Not observed | |
| | ***CCN*** | ***2.1*** | ***Observed*** | [36] |
| | | | | |
| $C_2P$ | CCP | 0.0 | Observed | [37] |
| | CPC | 85.3 | Not observed | |
| | | | | |
| $CHO^+$ | $HCO^+$ | 0.0 | Observed | [38] |
| | ***$HOC^+$*** | ***37.3*** | ***Observed*** | [39-40] |
| | | | | |
| $CHS^+$ | $HCS^+$ | 0.0 | Observed | [41] |
| | $HSC^+$ | 94.1 | Not observed | |
| | | | | |
| $CNO^-$ | $OCN^-$ | 0.0 | Observed | [42] |
| | $ONC^-$ | 17.1 | Not observed | |
| | | | | |
| CHS | HCS | 0.0 | Observed | [43] |
| | HSC | 41.5 | Observed | [43] |

## Isomers with 4 atoms

For the isomers with 4 atoms, 25 molecular species from 10 isomeric groups are examined. Table 5 contains the isomerization energies of these molecular species and their current astronomical status. Fourteen of these molecular species have been detected from different astronomical sources [45-63]. In the CHON, CHSN, and $C_3H$ groups where more than one isomer has been detected, the relative enthalpy of formation ranges from 3.1 to 67.4 kcal/mol. For the remaining 7 isomeric groups with only one known molecular species from each, the relative enthalpy of formation ranges from 21.6 to 49.8 kcal/mol. This range falls within that of the known species, thus, pointing to the possibility of detecting the less stable isomers of these groups that could be formed via the isomerization process.

Among the isomers with 4 atoms examined here, there are eight cyanide/isocyanide pairs. In two of these pairs, both cyanide and isocyanide have been detected with a relative enthalpy of formation in the range of 2.8 to 29.0 kcal/mol. For the other cyanide/isocyanide pairs where only the cyanides are observed, the relative enthalpy of formation ranges from 18.7 to 24.7 kcal/mol which is below the 29.0 kcal/mol relative enthalpy of formation calculated for the known cyanide/isocyanide pairs. Thus, isomerization remains a plausible mechanism for the formation of the less stable isocyanides from their corresponding cyanides which are probably more abundant.

The $C_2N_2$ isomeric group is the isoelectronic analogue of the $C_2NP$ isomeric group. Just as there is interesting chemistry between the O and S-containing interstellar molecular species; such also exists for the N and P-containing species, though it is not well explored as that of O and S. Over 80% of all the known interstellar and circumstellar molecules have been detected via their rotational spectral features because, at the low temperature of the ISM, rotational excited states are easily populated. $NC_2N$ (the most stable isomer of the $C_2N_2$ group) is microwave inactive meaning that it cannot be astronomically detected via radio astronomy. There are reasons to believe the presence and detectability of $NC_2N$; its protonated analogue, $NC_2NH^+$ has been detected [64]. NCCP, the isoelectronic analogue of $NC_2N$ is also known [62]. Both the protonated analogues of $NC_2N$ and its isoelectronic analogue that have been observed are likely to be less abundant than $NC_2N$ because of the reactive nature of the ion and the less cosmic abundance of P as compared to N. Thus, infrared astronomy of NCCN or radio astronomy of its isotopologues (which are microwave active) is likely to be successful.

The observed isomers in all the groups shown in Table 5 are also the most stable isomers.

Table 5: Isomerization enthalpies for isomers with 4 atoms

| Isomeric group | Isomers | Relative $\Delta_f H^O$ (kcal/mol) | Astronomical status | Reference |
|---|---|---|---|---|
| CHON | Isocyanic acid | 0.00 | Observed | [45] |
| | Cyanic acid | 29.0 | Observed | [46-47] |
| | Fulminic acid | 67.4 (0.0) | Observed | [48] |
| | Isofulminic acid | 86.1 (18.7) | Not observed | |
| | | | | |
| CHSN | HNCS | 0.0 | Observed | [49] |
| | HSCN | 11.2 | Observed | [50] |
| | HCNS | 40.6 (0.0) | Not observed | |
| | HSNC | 43.3 (2.8) | Not observed | |
| | | | | |
| $C_3H$ | c-$C_3H$ | 0.0 | Observed | [51] |
| | l-$C_3H$ | 3.1 | Observed | [52] |
| | | | | |
| $C_3N$ | l-$C_3N$ | 0.0 | Observed | [53-54] |
| | $C_2NC$ | 22.7 | Not observed | |
| | c-$C_3N$ | 28.8 | Not observed | |
| | | | | |
| $C_2NH$ | $HC_2N$ | 0.0 | Observed | [55] |
| | HCNC | 28.0 | Not observed | |
| | | | | |
| $C_3O$ | l-$C_3O$ | 0.0 | Observed | [56-57] |
| | c-$C_3O$ | 21.6 | Not observed | |
| | | | | |
| $C_3S$ | l-$C_3S$ | 0.0 | Observed | [58-60] |
| | c-$C_3S$ | 26.3 | Not observed | |
| | | | | |
| $SiC_3$ | c-$C_3Si$ | 0.0 | Observed | [61] |
| | l-$C_3Si$ | 49.8 | Not observed | |
| | | | | |
| $C_2NP$ | NCCP | 0.0 | Observed | [62] |
| | CNCP | 24.1 | Not observed | |
| | | | | |
| $C_2N_2$ | $NC_2N$ | 0.0 | Not observed | |
| | CNCN | 24.7 | Observed | [63] |

**Isomers with 5 atoms**

Isomerization energies for 20 molecular species from 6 isomeric groups are presented in Table 6 alongside their current astronomical status. Nine of these species are known astromolecules [65-73]. In the $C_3HN$ and $C_3H_2$ groups where more than one isomer has been detected, the isomerization energy ranges from 13.7 to 45.4 kcal/mol, and the isomerization energies for some of the isomers with five atoms whose most stable isomers have been detected fall within this range. These include; ethynol, $CH_2NN$, $NH_2NC$, $CH_2NC$, HCONC and c-$C_2NHO$. These species can as well be formed from their most stable isomers via the isomerization mechanism.

Again, isomerization appears to be a favourable route for the formation of isocyanides from their corresponding cyanides. As shown in Table 6, an isomerization enthalpy as high as 45.4 kcal/mol is noted for an isocyanide formation. This strongly supports the formation of other isocyanides from their corresponding cyanides. The barrier for other isocyanides whose corresponding cyanides have been detected ranges from 12.9 to 43.4 kcal/mol which is within the limit of the one that has been observed.

The fact that the most stable isomers are probably the most abundant and are easily detected in ISM is well demonstrated among these isomers. From table 6, all the observed isomers are the most stable ones in their respective isomeric groups as compared to the ones that have not been detected.

Table 6: Isomerization enthalpies for isomers with 5 atoms

| Isomeric group | Isomers | Relative $\Delta_f H^O$ (kcal/mol) | Astronomical status | Reference |
|---|---|---|---|---|
| $C_3HN$ | HCCCN | 0.0 | Observed | [65] |
| | HCCNC | 24.1 | Observed | [66] |
| | **HNCCC** | **45.4** | **Observed** | [67] |
| | CC(H)CN | 50.5 | Not observed | |
| | HCNCC | 72.6 | Not observed | |
| | | | | |
| $C_2H_2O$ | Ketene | 0.0 | Observed | [68] |
| | Ethynol | 38.8 | Not observed | |
| | Oxirene | 81.9 | Not observed | |
| | | | | |
| $C_3H_2$ | c-$C_3H_2$ | 0.0 | Observed | [69] |
| | $l$-$C_3H_2$ | 13.7 | Observed | [70] |
| | | | | |
| $N_2H_2C$ | $NH_2CN$ | 0.0 | Observed | [71] |
| | $CH_2NN$ | 26.3 | Not observed | |
| | $NH_2NC$ | 43.4 | Not observed | |
| | | | | |
| $C_2H_2N$ | $CH_2CN$ | 0.0 | Observed | [72] |
| | $CH_2NC$ | 22.2 | Not observed | |
| | | | | |
| $C_2HNO$ | CNCHO | 0.0 | Observed | [73] |
| | HCONC | 12.9 | Not observed | |
| | c-$C_2NHO$ | 30.2 | Not observed | |
| | HNCCO | 69.2 | Not observed | |
| | $HC_2NO$ | 83.3 | Not observed | |

**Isomers with 6 atoms**

The isomerization energies and the current astronomical status of the different isomeric species with 6 atoms investigated in this study are presented in Table 7. Figures 1 to 5 display some of the cyclic isomers highlighted in Table 7. One-third of all the molecular species presented in Table 7 have all been detected from several astronomical sources [74-85]. In the $C_2H_3N$ and $H_2C_3O$ groups where more than one isomer has been detected, the isomerization enthalpies range from 7.3 to 23.1 kcal/mol. Except for some of the cyclic isomers which are highly unstable, most of the unknown isomers (c-$C_5N$*, $C_4NC$, c-$C_5N$**, c-$C_5H^a$, $SiH_3NC$, $HC_3NC$, methylene ketene, hydroxymethylimine) have isomerization enthalpies in the range of the known isomers, suggesting their possible astronomical observation if they could be formed via isomerization as the ones that have been observed. Ketenimine for instance is reported to be formed from methyl cyanide via tautomerization (i.e., an isomerization pathway in which the migration of hydrogen atom from the methyl group to the nitrogen atom is accompanied by a rearrangement of bonding electrons) and this process is said to be driven by shock waves that provide the energy for both the formation and distribution of large interstellar species [80].

The $C_5H$ isomers; l-$C_5H$ and c-$C_5H^a$ are almost identical in energy. If the spectroscopic parameters of c-$C_5H^a$ (Figure 2a) are accurately probed, either experimentally or theoretically; this species (c-$C_5H$) could be detected in ISM. The four cyanide/isocyanide pairs among the isomers with 6 atoms presented in Table 7 have isomerization enthalpies in the range of 4.5 to 23.0 kcal/mol. Interestingly, the $C_2H_3N$ group where methyl cyanide and methyl isocyanide have been detected has the highest relative enthalpy of formation of 23.0 kcal/mol which implies the possible detection of the other isocyanides with lower barriers (4.5 to 20.7kcal/mol) assuming isomerization as their plausible formation route.

Methylene ketene and methyl ketene have been proposed as potential interstellar molecules for many reasons. The ketenes are found to be more stable than their corresponding isomers (Tables 6, 7, and 9); they are less affected by interstellar hydrogen bonding assuming surface formation processes; ketene and ketenyl radical ($H_2C_2O$ and $HC_2O$ respectively) are known interstellar molecules [64; 68].

Just as every known O-containing interstellar molecule points to the presence and the detectability of the S-analogue. For every known S-species, the O-analogue is not only present in detectable abundance, but it can also be said to have even been overdue for astronomical detection. $C_5O$ is the O-analogue of $C_5S$ that has been detected. Thus, it is an important potential interstellar molecule.

Table 7: Isomerization enthalpies for isomers with 6 atoms

| Isomeric group | Isomers | Relative $\Delta_f H^O$ (kcal/mol) | Astronomical status | Reference |
|---|---|---|---|---|
| $C_5N$ | l-$C_5N$ | 0.0 | Observed | [74] |
| | c-$C_5N$* | 17.4 | Not observed | |
| | $C_4NC$ | 20.7 | Not observed | |
| | c-$C_5N$** | 21.0 | Not observed | |
| | c-$C_5N$*** | 100.1 | Not observed | |
| | | | | |
| $C_5H$ | l-$C_5H$ | 0.0 | Observed | [75-77] |
| | ***c-$C_5H^a$*** | ***1.0*** | ***Not observed*** | |
| | c-$C_5H^b$ | 34.7 | Not observed | |
| | c-$C_5H^c$ | 60.8 | Not observed | |
| | | | | |
| $C_2H_3N$ | Methyl cyanide | 0.0 | Observed | [78] |
| | Methyl isocyanide | 23.0 | Observed | [79] |
| | **Ketenimine** | **23.1** | **Observed** | [80] |
| | Ethynamine | 42.3 | Not observed | |
| | 2H-azirine | 46.5 | Not observed | |
| | 1H-azirine | 80.8 | Not observed | |
| | | | | |
| $SiCH_3N$ | $SiH_3CN$ | 0.0 | Observed | [62] |
| | $SiH_3NC$ | 4.5 | Not observed | |
| | c-$SiCH_3N^d$ | 33.4 | Not observed | |
| | $H_2SiCNH$ | 46 | Not observed | |
| | c-$SiCH_3N^e$ | 50.2 | Not observed | |
| | $H_2NCSiH$ | 57.0 | Not observed | |
| | | | | |
| $C_4HN$ | $HC_4N$ | 0.0 | Observed | [81] |
| | $HC_3NC$ | 17.4 | Not observed | |
| | | | | |
| $H_2C_3O$ | Methylene ketene | 0.0 | ***Not observed*** | |
| | Propynal | 7.3 | Observed | [82] |
| | Cyclopropenone | 10.6 | observed | [83] |
| | | | | |
| $H_3CON$ | Formamide | 0.0 | Observed | [84] |
| | Hydroxymethylimine | 12.6 | Not observed | |
| | Nitrosomethane | 60.8 | Not observed | |
| | | | | |
| $C_5S$ | l-$C_5S$ | 0.0 | Observed | [62; 85] |
| | c-$C_5S^{\#}$ | 26.9 | Not observed | |
| | c-$C_5S^{\#\#}$ | 41.8 | Not observed | |
| | c-$C_5S^{\#\#\#}$ | 84.1 | Not observed | |
| | $C_4SC$ | 114.2 | Not observed | |
| | | | | |
| $C_5O$ | l-$C_5O$ | 0.0 | Not observed | |
| | c-$C_5O^{\sigma}$ | 28.6 | Not observed | |
| | c-$C_5O^{\sigma\sigma}$ | 46.3 | Not observed | |
| | c-$C_5O^{\sigma\sigma\sigma}$ | 84.5 | Not observed | |
| | $C_4OC$ | 114.3 | Not observed | |

*Figure 1a, **Figure 1b, ***Figure 1c; [a]Figure 2a, [b]Figure 2b, [c]Figure 2c; #Figure 3a, ##Figure 3b, ###Figure 3c; [d]Figure 4a, [e]Figure 4b; [σ]Figure 5a, [σσ]Figure 5b, [σσσ]Figure 5c.

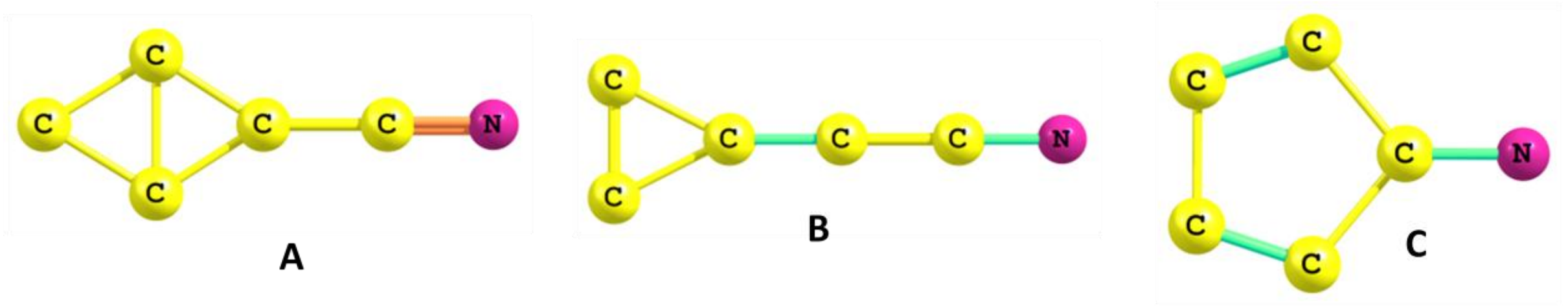

Figure 1: Optimized structures of cyclic $C_5N$ isomers.

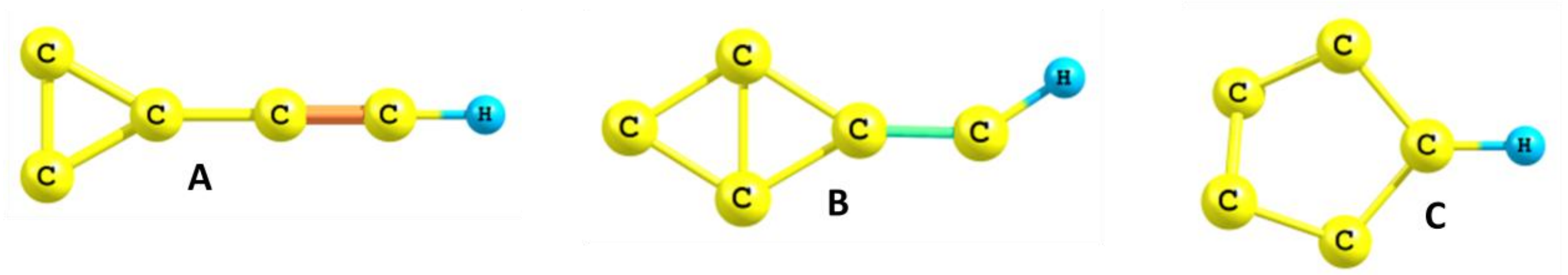

Figure 2: Optimized structures of cyclic $C_5H$ isomers.

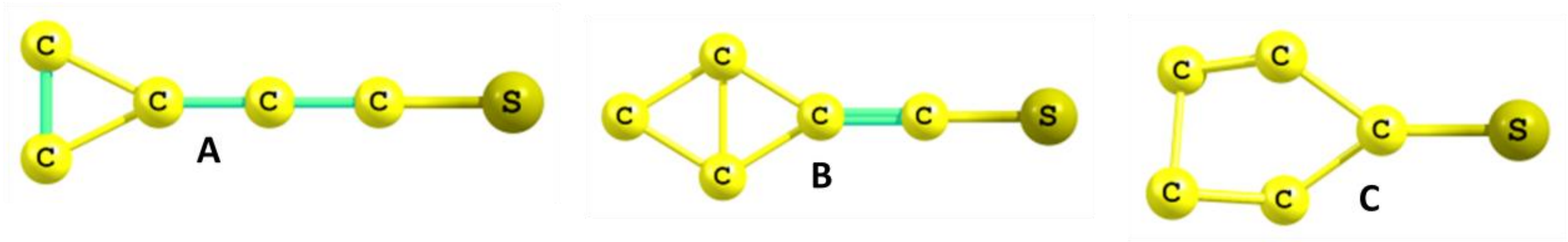

Figure 3: Optimized structures of cyclic $C_5S$ isomers.

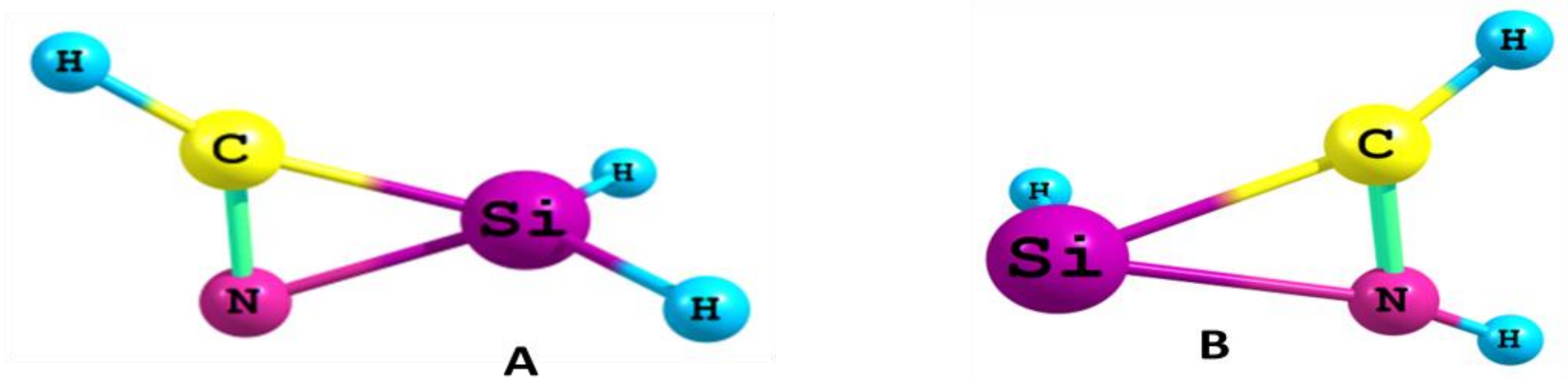

Figure 4: Optimized structures of cyclic $SiCH_3N$ isomers.

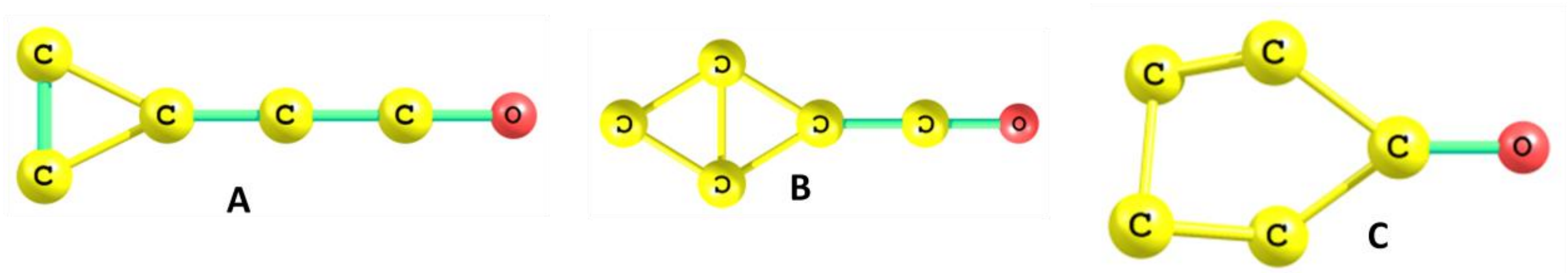

Figure 5: Optimized structures of cyclic $C_5O$ isomers.

**Isomers with 7 atoms**

Table 8 shows the isomerization energies and the current astronomical status of different isomeric species with 7 atoms investigated in this study. Of the 22 molecular species in the Table, 7 have been astronomically observed [86-93]. In the $H_4C_2O$ isomeric group where all the isomers have been detected, the isomerization energies range from 12.2 to 27.8 kcal/mol. All the non-observed isomers in the $H_3C_3N$ and $C_3H_4$ isomeric groups including, cyanomethanol, iminoacetaldehyde, and methyl cyanate (from $C_2H_3NO$ isomeric group) have isomerization enthalpies in the range calculated for the $H_4C_2O$ isomers which have all been detected. This points to the possibility of detecting these molecules in ISM with relative enthalpy of formation within those of the known interstellar molecules. There is no cyanide/isocyanide pair observed among these systems, however, the relative enthalpy of formation in the range of 19.9 to 27.3 kcal/mol is within the range of those noted for other observed cyanide/isocyanide pairs.

As in the previous examples, all the observed isomers are the most stable ones in their respective groups. In the $H_4C_2O$ isomeric group where all the stable isomers have been observed, the most stable isomer; acetaldehyde was first observed before the other isomers. As would be expected, the most stable isomer (acetaldehyde) is present in high abundance in the different astronomical sources where it has been detected as compared to the abundances of the less stable isomers in the same sources [94-96].

Table 8: Isomerization enthalpies for isomers with 7 atoms

| Isomeric group | Isomers | Relative $\Delta_f H^O$ (kcal/mol) | Astronomical status | Reference |
|---|---|---|---|---|
| $H_4C_2O$ | Acetaldehyde | 0.0 | Observed | [86-87] |
| | Vinyl alcohol (syn) | 12.2 | Observed | [88] |
| | Vinyl alcohol (anti) | 13.9 | Observed | [88] |
| | Ethylene oxide | 27.8 | Observed | [89] |
| | | | | |
| $H_3C_3N$ | Acrylonitrile | 0.0 | Observed | [90] |
| | Isocyanoethene | 19.9 | Not observed | |
| | | | | |
| $C_3H_4$ | $CH_3C_2H$ | 0.0 | Observed | [91] |
| | $H_2CCCH_2$ | 7.8 | Not observed | |
| | c-$C_3H_4$ | 23.6 | Not observed | |
| | | | | |
| $C_2H_3NO$ | Methyl isocyanate | 0.0 | Observed | [92-93] |
| | Cyanomethanol | 13.6 | Not observed | |
| | Iminoacetaldehyde | 20.1 | Not observed | |
| | Methyl cyanate | 27.3 | Not observed | |
| | 2-Aziridinone | 36.0 | Not observed | |
| | 2-Oxiranimine | 40.8 | Not observed | |
| | Methyl fulminic acid | 56.5 | Not observed | |
| | 2-Iminoethenol | 60.9 | Not observed | |
| | Nitrosoethene | 69.3 | Not observed | |
| | 2H-1,2-Oxazete | 79.8 | Not observed | |
| | Methyl isofulminic acid | 84.8 | Not observed | |
| | N-Hydroxyacetylenamin | 93.6 | Not observed | |
| | (Aminooxy)acetylene | 94.7 | Not observed | |

**Isomers with 8 atoms**

Table 9 contains the isomerization energies for 33 isomeric species from 7 groups containing eight atoms each. The astronomical statuses of these species are also shown in the table. Figure 6 highlights the cyclic isomers from the $C_2H_5N$ and $CH_4N_2O$ groups. As shown in Table 9, 11 of these species have been astronomically detected [97-109]. The isomerization energies of these species range from 3.1 to 50.0 kcal/mol. Except for 1,2-dioxetane, 1,3-dioxetane, and epoxyproprene, the isomerization energies for all the unknown isomers with 8 atoms fall within the range (of 13.0 to 46.9 kcal/mol) of the known isomers (3.1 to 50.0 kcal/mol). Thus, some of the unknown isomers in Table 9 could be formed from their most stable isomers (which are probably more abundant) via the isomerization process.

The three cyanide/isocyanide pairs among these isomers have isomerization enthalpies ranging from 19.0 to 25.6 kcal/mol. Though no cyanide/isocyanide pair has been detected among these molecular species, the relative enthalpy of formation is within the range of those where the pairs have been detected. $CH_3CCNC$ and $H_2NCH_2NC$ could be formed from their corresponding cyanides that have been detected, thus, they could also be detected provided their spectroscopic parameters are accurately known. Methyl ketene, the most stable isomer of the $H_4C_3O$ group is yet to be astronomically detected. The reasons for its presence and detectability in ISM are the same as discussed for methylene ketene (among the isomers with 6 atoms).

Table 9: Isomerization enthalpies for isomers with 8 atoms

| Isomeric group | Isomers | Relative $\Delta_f H^O$ (kcal/mol) | Astronomical status | References |
|---|---|---|---|---|
| $C_4H_3N$ | $CH_3CCCN$ | 0.0 | Observed | [97] |
| | $CH_2CCHCN$ | 3.1(0.0) | Observed | [98-99] |
| | $HCCCH_2CN$ | 13.0 | Not observed | |
| | $CH_3CCNC$ | 25.6 | Not observed | |
| | $CH_2CCHNC$ | 25.7 (22.6) | Not observed | |
| | | | | |
| $C_2H_4N_2$ | $H_2NCH_2CN$ | 0.0 | Observed | [100] |
| | $H_2NCH_2NC$ | 19.0 | Not observed | |
| | | | | |
| $C_2H_4O_2$ | Acetic acid | 0.0 | Observed | [101] |
| | Methylformate | 17.7 | Observed | [102-103] |
| | Glycolaldehyde | 33.2 | Observed | [104] |
| | 1,3-dioxetane | 52.9 | Not observed | |
| | 1,2-dioxetane | 103.0 | Not observed | |
| | | | | |
| $H_4C_3O$ | Methyl ketene | 0.0 | Not observed | |
| | Propenal | 2.3 | Observed | [105] |
| | Cyclopropanone | 17.3 | Not observed | |
| | Propynol | 30.8 | Not observed | |
| | Propargyl alcohol | 35.2 | Not observed | |
| | Methoxy ethyne | 42.0 | Not observed | |
| | 1-cyclopropenol | 44.0 | Not observed | |
| | 2-cyclopropenol | 45.5 | Not observed | |
| | Epoxypropene | 65.8 | Not observed | |
| | | | | |
| $H_2C_6$ | $HC_6H$ | 0.0 | Observed | [106] |
| | ***$H_2C_6$*** | ***50.0*** | ***Observed*** | [107] |
| | | | | |
| $C_2H_5N$ | $CH_3CHNH$ | 0.0 | Observed | [108] |
| | $H_2CCHNH_2$ | 2.8 | Not observed | |
| | CH3NCH2 | 8.1 | Not observed | |
| | c-$C_2H_5N$[m] | 19.8 | Not observed | |
| | | | | |
| $CH_4N_2O$ | $H_2NCONH_2$ | 0.0 | Observed | [109] |
| | $H_2NNHCHO$ | 11.0 | Not observed | |
| | $HN_2CH_2OH$ | 35.0 | Not observed | |
| | c- $CH_4N_2O$[n] | 42.4 | Not observed | |
| | $CH_3NHNO$ | 43.4 | Not observed | |
| | $H_2NCHNHO$ | 46.9 | Not observed | |

[m]Figure 5a; [n]Figure 5b.

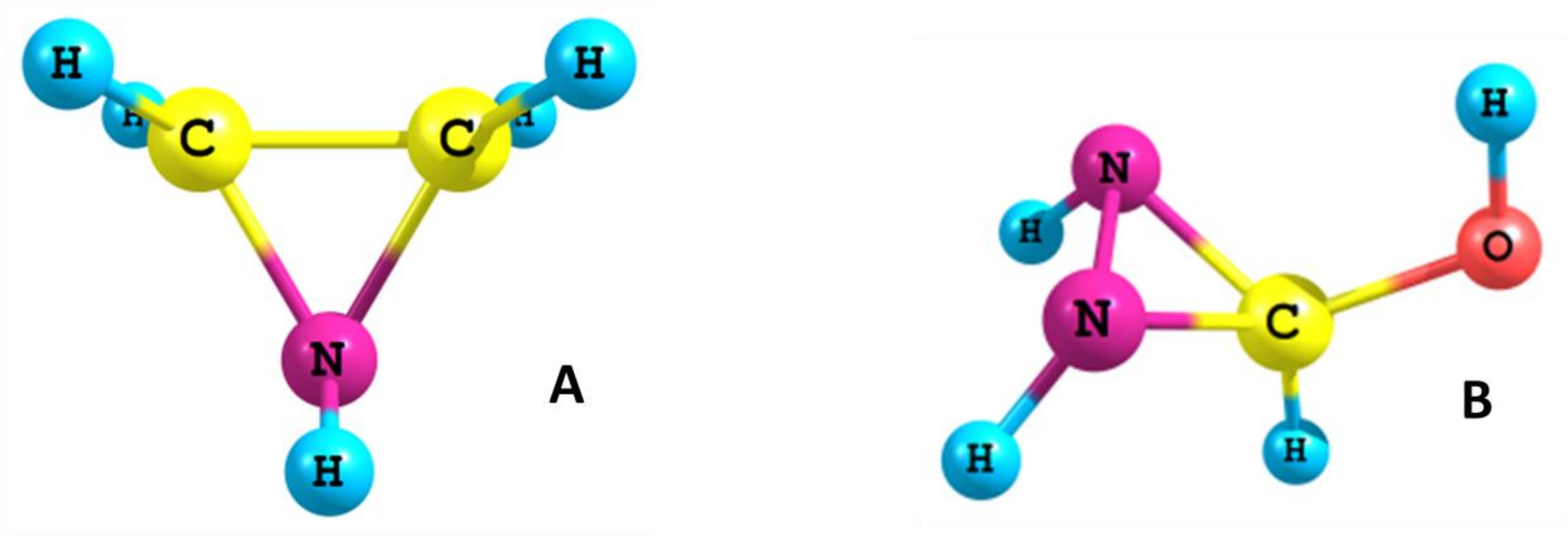

Figure 6: Optimized structures of cyclic $C_2H_5N$ (A) and $CH_4N_2O$ (B) isomers.

**Isomers with 9 atoms**

In Table 10, 28 isomeric species from 7 groups are presented with their isomerization enthalpies and current astronomical status. Eight of these species are known interstellar molecular species [110-120]. Figure 7 shows some of the cyclic molecules with 9 atoms. Among the species shown in Table 10, only ethanol and dimethyl ether from the $C_2H_6O$ group have been observed in more than one isomeric form with a relative enthalpy of formation of 12.6 kcal/mol. $CH_3SCH_3$, the S-analogue of dimethyl ether remains a potential candidate for astronomical detection for many reasons; the well-established chemistry of S and O-containing interstellar molecules, the high abundance of $CH_3CH_2SH$ (the most stable isomer of the group which can isomerize to $CH_3SCH_3$) and the low relative enthalpy of formation (1.4 kcal/mol) as compared to that of dimethyl ether (12.6 kcal/mol). Assuming accurate spectroscopic parameters are available, interstellar $CH_3SCH_3$ will soon become a reality.

Isomerization remains one of the plausible formation routes for the formation of the less stable isomers of the different groups here whose most stable isomers have been detected. The isomerization enthalpies for the two cyanide/isocyanide pairs range from 20.8 to 27.1 kcal/mol. Though none of these pairs has been detected, the possibility of their formation from their corresponding cyanides (via isomerization) which could lead to their successful detection cannot be ruled out. $HC_6NC$ is the highest member of the $HC_{2n}NC$ linear chains with experimentally measured rotational transitions that can be used for its astronomical search [121]. However, a recent study using a combined experimental and theoretical approach has provided accurate rotational constants for higher members of the $HC_{2n}NC$ linear chains [6]. HNC remains the only member of this series that has been detected in ISM [122].

Table 10: Isomerization enthalpies for isomers with 9 atoms

| Isomeric group | Isomers | Relative $\Delta_f H^O$ (kcal/mol) | Astronomical status | References |
|---|---|---|---|---|
| $C_2H_6O$ | Ethanol | 0.0 | Observed | [110-111] |
| | Dimethyl ether | 12.6 | Observed | [112] |
| | | | | |
| $C_2H_6S$ | $CH_3CH_2SH$ | 0.0 | Observed | [113] |
| | $CH_3SCH_3$ | 1.4 | Not observed | |
| | | | | |
| $C_2H_5ON$ | Acetamide | 0.0 | Observed | [114] |
| | N-methylformamide | 9.7 | Not observed | |
| | Nitrosoethane | 64.7 | Not observed | |
| | 1-aziridnol | 77.9 | Not observed | |
| | Cyanoethoxyamide | 134.4 | Not observed | |
| | | | | |
| $C_3H_5N$ | Cyanoethane | 0.0 | Observed | [115] |
| | Isocyanoethane | 20.8 | Not observed | |
| | Propylenimine | 22.0 | Not observed | |
| | 2-propen-1-imine | 25.3 | Not observed | |
| | N-methylene ethenamine | 26.0 | Not observed | |
| | Azastene | 32.2 | Not observed | |
| | Cyclopropanimine | 37.2 | Not observed | |
| | Methylene azaridine | 43.8 | Not observed | |
| | Propargylamine | 45.1 | Not observed | |
| | 1-azabicyclo(1.1.0)butane | 53.5 | Not observed | |
| | | | | |
| $C_5H_4$ | $CH_3C_4H$ | 0.0 | Observed | [116] |
| | $H_2C_3HC_2H$ | 5.2 | Not observed | |
| | $H_2C_5H_2$ | 8.1 | Not observed | |
| | c- $C_5H_4$[x] | 26.1 | Not observed | |
| | c- $C_5H_4$[y] | 31.6 | Not observed | |
| | | | | |
| $C_3H_6$ | $CH_3CHCH_2$ | 0.0 | Observed | [117] |
| | c- $C_3H_6$[z] | 8.2 | Not observed | |
| | | | | |
| $C_7HN$ | $HC_7N$ | 0.0 | Observed | [118-120] |
| | $HC_6NC$ | 27.1 | Not observed | |

[x]Figure 6a; [y]Figure 6b; [z]Figure 6c.

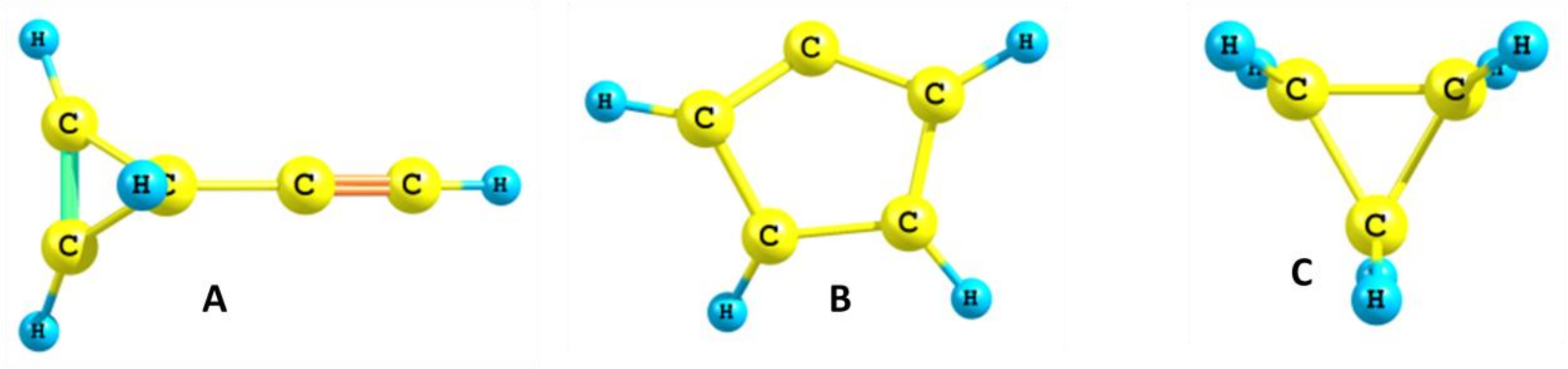

Figure 7: Optimized structures of cyclic $C_5H_4$ (A and B) and $C_3H_6$ (C) isomers.

## Isomers with 10 atoms

Isomerization energies and astronomical statuses for 16 molecular species from three isomeric groups comprising 10 atoms each are presented in Table 11. Of these three groups, only the $C_3H_6O$ group contains isomers (propanone and propanal) that have been detected in more than one isomeric form with a relative enthalpy of formation of 8.1 kcal/mol [123-127]. $CH_3(CC)_2NC$ is the isocyanide analogue of $CH_3(CC)_2CN$ that has been detected. It has a relative enthalpy of formation of 25.3 kcal/mol which is within the range of the known cyanide/isocyanide pairs. However, there is little or no information regarding the rotational spectrum of $CH_3(CC)_2NC$ that could warrant its astronomical search. Thus, the availability of accurate spectroscopic parameters for this species remains the starting point in the steps toward its astronomical detection.

1,1-ethanediol; the most stable isomer of the $C_2H_6O_2$ group has not been detected whereas ethylene glycol, the next stable isomer of the group with an isomerization enthalpy of 6.1kcal/mol has been detected in good abundance. The delayed astronomical detection of 1,1-ethanediol is directly linked to a lack of spectroscopic parameters for this molecule. The rotational spectrum of 1,1-ethanediol is yet to be probed either experimentally or theoretically. Once this is done, this molecule could be successfully observed.

Table 11: Isomerization enthalpies for isomers with 10 atoms

| Isomeric group | Isomers | Relative $\Delta_f H^O$ (kcal/mol) | Astronomical status | References |
|---|---|---|---|---|
| $C_2H_6O_2$ | 1,1-Ethanediol | 0.0 | Not observed | |
| | Ethylene glycol | 6.1 | Observed | [123] |
| | Methoxy methanol | 15.9 | Not observed | |
| | Ethyl hydroperoxide | 51.9 | Not observed | |
| | Dimethane peroxide | 57.7 | Not observed | |
| | | | | |
| $C_3H_6O$ | Propanone | 0.0 | Observed | [124-125] |
| | Propanal | 8.1 | Observed | [126] |
| | Propen-2-ol | 13.8 | Not observed | |
| | 1-propen-1-ol | 18.6 | Not observed | |
| | Methoxy ethene | 25.6 | Not observed | |
| | 2-propene-1-ol | 27.1 | Not observed | |
| | 1,2-epoxypropane | 30.0 | Not observed | |
| | Cyclopropanol | 31.2 | Not observed | |
| | Oxetane | 33.2 | Not observed | |
| | | | | |
| $C_6H_3N$ | $CH_3(CC)_2CN$ | 0.0 | Observed | [127] |
| | $CH_3(CC)_2NC$ | 25.3 | Not observed | |

**Isomers with 11 atoms**

There are currently seven known interstellar molecules ($HC_9N$, $CH_3C_6H$, ethyl formate, methyl acetate, hydroxyacetone, cyclopentadiene and ethanolamine) containing 11 atoms [128-134]. In Table 12, the 9 isomeric species from two groups containing 11 atoms are shown with their isomerization enthalpies and current astronomical status. For the known isomers in the $C_3H_6O_2$ group, the isomerization energy ranges from 11.8 to 14.3 kcal/mol. The non-detection or delayed detection of propanoic acid (the most stable isomer of the group) has been traced to the effect of interstellar hydrogen bonding on the surface of the dust grains which reduces the overall abundance of this species in the gas phase, thus, influencing its successful astronomical observation. $HC_9N$ is the second largest member of the cyanopolyyne chain that has been detected in ISM. As seen in other cyanide/isocyanide pairs, the isomerization of $HC_9N$ to $HC_8NC$ with the relative enthalpy of formation of 27.2 is achievable. An accurate rotational constant for $HC_8NC$ is now available from a recent study [6; 17]. Thus, $HC_8NC$ could be astronomically searched.

Table 12: Isomerization enthalpies for isomers with 11 atoms

| Isomeric group | Isomers | Relative $\Delta_f H^O$ (kcal/mol) | Astronomical status | References |
|---|---|---|---|---|
| $C_3H_6O_2$ | Propanoic acid | 0.0 | Not observed | |
| | Ethylformate | 11.8 | Observed | [128] |
| | Methyl acetate | 14.3 | Observed | [129] |
| | Lactaldehyde | 28.1 | Not observed | |
| | Dioxolane | 36.1 | Not observed | |
| | Glycidol | 52.1 | Not observed | |
| | Dimethyldioxirane | 81.7 | Not observed | |
| | | | | |
| $C_9HN$ | $HC_9N$ | 0.0 | Observed | [130] |
| | $HC_8NC$ | 27.2 | Not observed | |

**Isomers with 12 atoms**

Only very few cyclic molecules are known among the interstellar molecular species [17; 20-21. From the different isomeric groups considered in this study, it is clear that the cyclic isomers are found to be among the less stable isomers in their respective groups as compared to their corresponding straight-chain analogues. This low stability could affect their interstellar abundance and influences their astronomical observation. The $C_6H_6$ isomeric group is one of the few groups where the cyclic molecules are found to be the most stable species. Apart from cyanocyclopentadiene and 2-cyanocyclopentadiene that have been recently observed in the interstellar medium [135-136], the other four known interstellar molecules with 12 atoms, [137-140] their corresponding isomers, and isomerization enthalpies are presented in Table 13.

The only known branched-chain interstellar molecule; isopropyl cyanide is almost equivalent in energy to its linear chain analogue; propyl cyanide with a relative enthalpy of formation of 0.4kcal/mol. Propan-2-ol and propanol (the two most stable isomers of the $C_3H_6O$ group) are yet to be astronomically observed. These species have also been shown to be strongly affected by interstellar hydrogen bonding as compared to ethyl methyl ether that has been detected [6; 18-21]. The stability of ethyl methyl ether is also enhanced by the stabilizing effect of the two alkyl substituents while propanol and propan-2-ol have only one alkyl substituent each.

Table 13: Isomerization enthalpies for isomers with 12 atoms

| Isomeric group | Isomers | Relative $\Delta_f H^O$ (kcal/mol) | Astronomical status | References |
|---|---|---|---|---|
| $C_6H_6$ | Benzene | 0.0 | Detected | [137] |
| | Fulvene | 34.2 | Not detected | |
| | 3,4-dimethylenecyclopropene | 62.0 | Not detected | |
| | 1,5-hexadiene-3-yne | 62.3 | Not detected | |
| | 2,4-hexadiyne | 65.2 | Not detected | |
| | 1,2,3,4-hexateraene | 71.7 | Not detected | |
| | 1,3-hexadiyne | 73.2 | Not detected | |
| | Trimethylenecyclopropane | 79.8 | Not detected | |
| | 1,4-hexadiyne | 82.3 | Not detected | |
| | 1,5-hexadiyne | 85.8 | Not detected | |
| | | | | |
| $C_3H_8O$ | Propan-2-ol | 0.0 | Not observed | |
| | Propanol | 3.7 | Not observed | |
| | Ethyl methyl ether | 8.2 | Observed | [139] |
| | | | | |
| $C_4H_7N$ | Isopropyl cyanide | 0.0 | Observed | [138] |
| | Propyl cyanide | 0.4 | Observed | [140] |
| | 2-isocyanopropane | 18.2 | Not observed | |
| | 2-aminobutadiene | 18.2 | Not observed | |
| | 3-pyrroline | 20.2 | Not observed | |
| | 2,2-dimethylethylenimine | 20.5 | Not observed | |
| | But-1-en-1-imine | 22.6 | Not observed | |
| | 2,3-butadiene-1-amine | 38.1 | Not observed | |
| | N-vinylazaridine | 40.1 | Not observed | |
| | N-methyl-1-propyn-1-amine | 41.4 | Not observed | |
| | 3-butyn-1-amine | 43.2 | Not observed | |
| | N-methyl propargylamine | 49.3 | Not observed | |
| | 2-azabicyclo(2.1.0)pentane | 51.9 | Not observed | |

**Potential Interstellar Molecules**

Knowing the right candidates for astronomical searches is vital in reducing the number of unsuccessful astronomical searches considering the time, energy and resources involved in these projects. From the present study, few molecular species have been identified as potential interstellar molecules. These are briefly summarized below:

**Isocyanomethylidyne, CNC**: This is the isocyanide analogue of $C_2N$ that has been detected [141]. CNC is found to be more stable than its cyanide analogue. Table 14 contains the relative energies of CNC and CCN from different quantum chemical calculation methods while Figure 8 shows the optimized geometries of these isomers at the MP2(full)/6-311++G** level with the bond angle in degrees and bond distance in angstroms. As discussed under the isomers with 3 atoms, all the observed species are the most stable ones and these stable species are also found to be more abundant. Where both species have been detected, the most stable is reported to present in high abundance than the less stable. For instance, HCN is found to be more abundant than HNC in different molecular clouds and this is also the case for the MgNC/MgCN abundance ratio measured in the asymptotic giant branch (AGM) stars [142-145]. CNC is more stable than CCN which implies that CNC should be present in high abundance in ISM than CCN that has been detected. Thus, CNC remains a potential candidate for astronomical detection. CNC is microwave inactive with a zero-dipole moment; thus, infrared astronomy remains the best approach for its astronomical observation.

Table 14: Relative energies of CNC and CCN

| Method | Relative energy (kcal/mol) | |
|---|---|---|
| | CNC | CCN |
| CCSD/6-311++G** | 0.0 | 1.5 |
| MP2(full)/6-311++G** | 0.0 | 6.2 |
| B3LYP/6-311++G** | 0.0 | 2.0 |
| G4 | 0.0 | 3.7 |

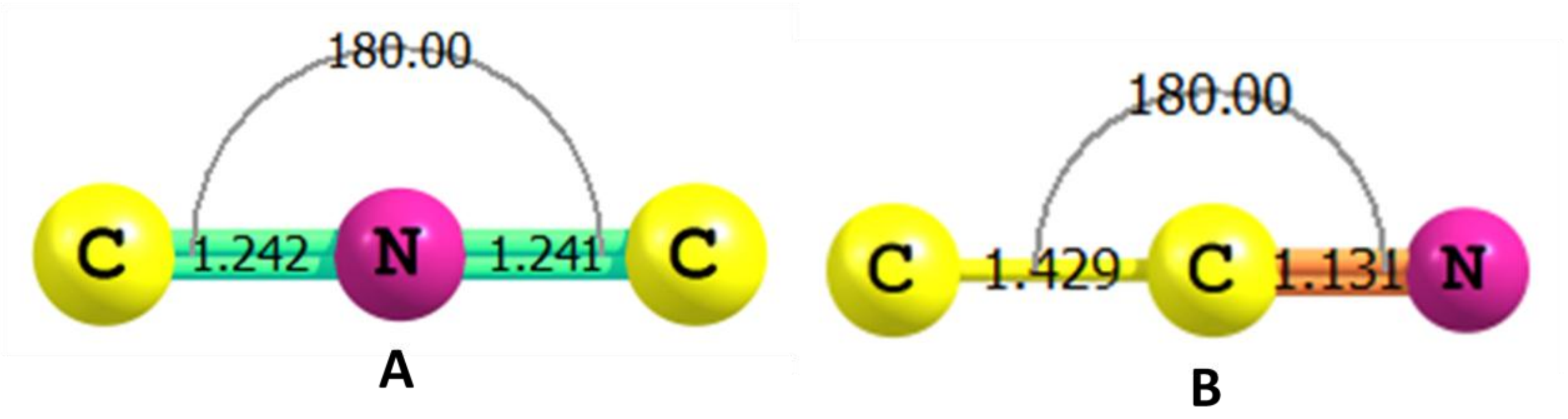

Figure 8: Optimized geometries of CNC and CCN at MP2(full)/6-311++G** level.

**NCCN**: The rationale for the choice of NCCN as a potential interstellar molecule is well discussed under the isomers with 4 atoms.

**c-$C_5$H**: With the limited number of known cyclic interstellar molecules, it is always exciting finding rings that are as stable as their corresponding chains. c-$C_5$H (Figure 2a) is the cyclic analogue of $C_5$H that has been detected. From Table 7, the relative enthalpy of formation between the linear chain and this cyclic analogue is just 1 kcal/mol. Table 15 shows the relative energies of these species at different quantum chemical calculation methods. While the MP2 method predicts the cyclic isomer to be more stable than the linear, others methods predict the reverse but the magnitude of the difference in energy is much at the MP2 level as compared to other methods. Figure 9 displays the optimized structures of these species. As shown in Table 15, c-$C_5$H is microwave active with a very good dipole moment making its astronomical searches in the radio frequency possible. If the spectroscopic parameters of c-C5H can be accurately probed, either experimentally or theoretically, the possibility of its astronomical observation is very high.

Table 15: Relative energies of linear and cyclic stable isomers of $C_5$H

| Method | Relative energy (kcal/mol) | | Dipole moment (Debye) | |
|---|---|---|---|---|
| | c-$C_5$H | l-$C_5$H | c-$C_5$H | l-$C_5$H |
| MP2(full)/6-311++G** | 0.0 | 18.7 | 3.4 | 4.2 |
| MP2/aug-cc-pVTZ | 0.0 | 17.5 | 3.4 | 4.2 |
| G4 | 10.9 | 0.0 | 3.7 | 4.6 |
| CCSD/6-311++G** | 3.0 | 0.0 | 3.4 | 4.6 |
| B3LYP/6-311++G** | 10.4 | 0.0 | 3.7 | 5.2 |
| B3LYP/aug-cc-pVTZ | 10.3 | 0.0 | 3.7 | 5.2 |

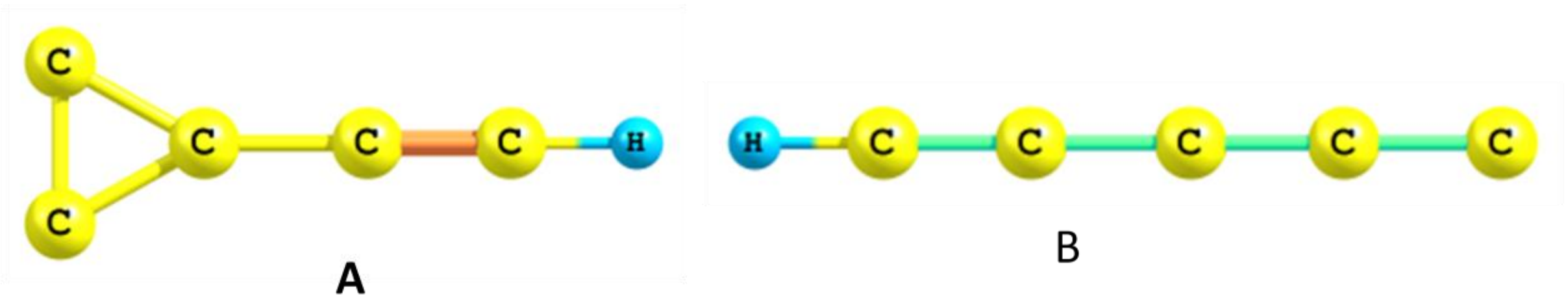

Figure 9: Optimized structures of $C_5$H most stable isomers.

**Methylene and Methyl Ketenes**: As highlighted in the text, the ketenes are found to be more stable than their corresponding isomers in all the isomeric groups examined in this study (Tables 6, 7 and 9) and as such be present in detectable amounts in ISM. Ketenes have also been shown to be less affected by hydrogen bonding on the surface of the interstellar dust grains as compared to their corresponding isomers. Also, the fact that ketene ($H_2C_2O$), ketenyl radical ($HC_2O$) and isomers of ketenes (like propynal, cyclopropenone, and propenal) are known interstellar molecular species further supports the presence and detectability of methyl ketene and methylene ketene in ISM. Thus, methyl ketene and methylene kenetne remain potential interstellar molecules pending their astronomical searches. No unsuccessful astronomical search has been reported for any of these ketenes.

**$C_5O$**: This is the oxygen analogue of $C_5S$ which is a known interstellar molecule. Without any exception, an interstellar O-containing molecular species is more abundant than its S-analogue. This could simply be traced to the cosmic abundance of O and S. Thus, for every known S-species, the O-analogue is not only present in detectable abundance, but it can also be said to have even been overdue for astronomical detection because for sure the O-species are more abundant than their S-analogue and as such could be detected with less difficulty as compared to its S-analogue. The column density of 2.14x $10^{12}cm^{-2}$ reported for $C_5S$ suggests the-high abundance of $C_5O$ in ISM [62; 85]. The microwave spectrum of $C_5O$ that could guide its successful astronomical observation is available [146]. Interstellar $C_5O$ is just a matter of time.

**$CH_3SCH_3$**: This is the S-analogue of dimethyl ether; a known interstellar molecule. The interstellar chemistry of S- and O-containing species is well established. Every known O-containing interstellar molecule points to the presence and detectability of the S-analogue. As earlier mentioned, except where the effect of interstellar hydrogen bonding dominates, the ratio of an interstellar sulphur molecule to its oxygen analogue is close to the cosmic S/O ratio. Assuming dimethyl ether is formed from ethanol via isomerization with a barrier of 12.6 kcal/mol, then the isomerization of $C_2H_5SH$ to $CH_3SCH_3$ is a much more feasible process with a relative enthalpy of formation of just 1.4 kcal/mol (nine times lower than that of $CH_3OCH_3$) compared to that of the O-analogue. Thus, because of the unique chemistry of S- and O-containing interstellar molecules and the low relative enthalpy of formation, $CH_3SCH_3$ is considered a potential candidate for astronomical observation assuming its spectroscopic parameters are accurately known.

**1,1-Ethanediol, Propanoic Acid, Propan-2-ol, and Propanol**: These molecules are found to be the most stable isomers in their respective groups (Tables 11, 12, and 13). Whereas their less stable isomers have been astronomically detected; successful detection of these most stable isomers is highly feasible. The spectroscopic parameters for propanoic acid, propan-2-ol, and propanol are known but those of 1,1-ethanediol are yet to come by. Though the delayed astronomical observations of some of these species have been linked to the effect of interstellar hydrogen bonding; there are high chances for their successful astronomical detection.

**Conclusion**: An extensive investigation of the isomerization energies of 246 molecular species from 65 isomeric groups is reported in this study. From the results, isomerization is found to be one of the plausible mechanisms for the formation of molecules in ISM. The high abundances of the most stable isomers coupled with the energy sources in ISM drive the isomerization process even for barriers as high as 67.4 kcal/mol. However, the chemical reaction pathways can also influence the final abundance of any species in the ISM. Specifically, the isomerization process is found to be very effective in converting cyanides to their corresponding isocyanides and vice versa. Thus, for every cyanide or isocyanide, the corresponding isocyanide or cyanide could be synthesized via isomerization. Also, from the results, the following potential interstellar molecules; CNC, NCCN, c-$C_5H$, methylene ketene, methyl Ketene, $CH_3SCH_3$, $C_5O$, 1,1-ethanediol, propanoic acid, propan-2-ol, and propanol are highlighted and discussed

**Acknowledgement:** EEE acknowledges a research fellowship from the Indian Institute of Science, Bangalore, and fruitful discussions with Prof. E. Arunan of the same Institute. The visiting position from the Indian Institute of Technology Bombay where the final part of this work was done is also acknowledged.